\documentclass[12pt]{spieman}  
\usepackage{amsmath,amsfonts,amssymb}
\usepackage{graphicx}
\usepackage{setspace}
\usepackage{tocloft}
\usepackage{epstopdf}
\usepackage{color}
\usepackage{tabularx}
\usepackage{multirow}
\usepackage{booktabs}
\usepackage{array}
\usepackage{changepage}

\title{Segmentation-Free X-ray Energy Spectrum Estimation for Computed Tomography Using Dual-Energy Material Decomposition}

\author[a,b]{Wei Zhao}
\author[b]{Lei Xing}
\author[a]{Qiude Zhang}
\author[a,*]{Qingguo Xie}
\author[c,*]{Tianye Niu}
\affil[a]{Huazhong University of Science and Technology, Department of Biomedical Engineering, 1037 Luoyu Rd, Wuhan, China, 430074}
\affil[b]{Stanford University, Department of Radiation Oncology, 875 Blake Wilbur Drive, Stanford, USA, 94305}
\affil[c]{Zhejiang University, School of Medicine, Sir Run Run Shaw Hospital and Institute of Translational Medicine, 3 Qingchun E Rd, Hangzhou, China, 310016}

\cftpagenumbersoff{figure}
\cftpagenumbersoff{table}
\begin{document}
\maketitle

\begin{abstract}
X-ray energy spectrum plays an essential role in computed tomography (CT) imaging and related tasks. Due to the high photon flux of clinical CT scanners, most of spectrum estimation methods are indirect and usually suffered from various limitations. In this study, we aim to provide a segmentation-free indirect transmission measurement-based energy spectrum estimation method using dual-energy material decomposition. The general principle of the method is to minimize the quadratic error between the polychromatic forward projection and the raw projection to calibrate a set of unknown weights which are used to express the unknown spectrum together with a set of model spectra. The polychromatic forward projection is performed using material-specific images which are obtained using dual-energy material decomposition. The algorithm has been evaluated using numerical simulations, experimental phantom data as well as realistic patient data. The results show the estimated spectrum matches the reference spectrum quite well and the method is robust. Extensive studies suggest the method provides accurate estimate of the CT spectrum without dedicated physical phantom and prolonged work flow. This paper may be attractive for CT dose calculations, artifacts reduction, polychromatic image reconstruction and other spectrum-involved CT applications.
\end{abstract}


{\noindent \footnotesize\textbf{*}Qingguo Xie,  \linkable{qgxie@hust.edu.cn} }\\
{\noindent \footnotesize\textbf{*}Tianye Niu,  \linkable{tyniu@zju.edu.cn} }

\begin{spacing}{2}   

\section{Introduction}
\label{sect:intro}  
X-ray computed tomography (CT) uses polychromatic x-ray photons to scan objects from different view angles. The energy spectrum of the x-ray photons determines the reconstructed image value, which is directly related to the application of CT scanners. Essentially, the spectrum plays \textcolor{black}{a} very important role in dose calculation\cite{demarco2005}, polychromatic image reconstruction\cite{elbakri2002}, artifacts reduction~\cite{nuyts2013,zhao2016}, spectral CT\cite{long2014,xi2015,zhao2015extended}, and \textcolor{black}{high contrast CT imaging\cite{zhao2015energy}}. To obtain the spectrum, a natural solution is to directly measure the energy of x-ray photons using energy-resolved detectors. However, in order to acquire high-quality diagnostic CT images, the x-ray photons flux of a CT scanner in clinical application is usually quite high (can exceed 1000 Mcps/mm$^2$\cite{taguchi2009}). For comparison, a typical energy-resolved detector is only 10 Mcps/mm$^2$\cite{taguchi2013}, thus it is not easy to directly measure the energy spectrum of the CT scanner using the energy-resolved detectors, as the detector pile-up effect limits the maximum count rate. Instead, spectrum calibration often employs indirect methods, including Compton-scattering measurement\cite{yaffe1976,matscheko1987,gallardo2004}, Monte Carlo (MC) simulation\cite{llovet2003,ay2004,mainegra2006,bazalova2007,miceli2007a}, empirical or semi-empirical physical models\cite{birch1979,tucker1991,boone1997,Hernandez2014} and transmission measurements\cite{archer1982,hassler1998,waggener1999,lin2014,chang2016l,perkhounkov2016x}.

Compton-scattering method aims to significantly reduce the incident photon flux with a factor of $10^5-10^6$. With this method, the scattered spectrum can be measured using a high purity germanium (HPGe) detector\cite{hammersberg1998,wilkinson2001} or a cadmium telluride detector\cite{maeda2005,duisterwinkel2015}. The incident spectrum is then reconstructed using the scattered spectrum and the corresponding scattering angle. However, the accuracy of spectrum estimated using this method may be suffered from the absorption in the scatterer and the limited absorption efficiency of the detector; thus careful system calibration is needed before yielding the final spectrum.

MC simulation can be easily used to generate the spectrum of an x-ray tube when its specifications (such as target angle and target material) are well known. In this case, the x-ray tube is modelled using a MC toolkit (such as Geant4\cite{allison2006}) with its specifications exactly match with that in the realistic application. Monochromatic electrons are emitted from an electron gun and accelerated to hit the target. Both bremsstrahlung and characteristic x-ray photons are generated and filtered by the inherent filtration material. Since most of the MC toolkits provide comprehensive physics modelling and are well validated\cite{taschereau2006,miceli2007,guthoff2012}, spectrum generated in this way can precisely match the real spectrum. However, the x-ray tubes are usually proprietary and it is difficult to obtain the exact tube specifications, especially for the tube filtration which is very important for spectrum simulation.

Spectral modelling calculates x-ray spectrum using mathematical models. These models are based on physical processes including bremsstrahlung and characteristic x-ray production. To yield \textcolor{black}{a} more accurate spectrum, \textcolor{black}{a} refined model that \textcolor{black}{accounts} self-absorption of the x-rays within the target was also introduced. Nowadays, the popular spectral modelling methods usually adopt an empirical or semisatirical approach that fits a parameterized model based on previously measured or simulated spectra\cite{birch1979,tucker1991,Hernandez2014}.

Transmission measurement is another methodology for spectrum estimation using projection data of a calibration phantom\cite{zhang2007,duan2011,lee2015}. The rationale of this method based on the fact that harder spectrum yields less attenuated projection while softer spectrum yields more attenuated projection. The method formulates the polychromatic forward projection equation as a discrete linear system in which the energy bins of the spectrum are described as unknown variables~\cite{sidky2005}, and then solve the linear system to obtain each bin contents of spectrum.

The accuracy of aforementioned methods is usually suffered from various limitations. For example, low-energy tailing yielded by hole trapping effect and environment conditions (such as low temperature requirement) may affect the spectrum measured using energy-resolved detectors\cite{koenig2012}. Attenuation and scattering (e.g., Rayleigh and multiple Compton) in the material of the scatterer of the Compton-scattering measurement need to be carefully considered\cite{duisterwinkel2015}. Transmission measurements based on step or wedge phantom requires dedicated hardware or workflow. The recently proposed indirect transmission measurements (ITM)\cite{zhao2015} needs at least the segmentation of one material class. When noise or artifacts are present in the reconstructed image, it causes incorrect material segmentation and yields inferior estimate of the spectrum.

Dual-energy CT (DECT) scanners have been widely used in realistic applications. Although x-ray spectrum is helpful for DECT material decomposition, especially for that in projection domain, accurate estimation of the high- and low- energy x-ray spectra is not a trivial task. Compared to projection domain material decomposition, image domain material decomposition does not suffer from inconsistent rays issue and is more convenient in clinical applications as it is performed on CT images\cite{maass2009,niu2014,mendonca2014}. In this case, x-ray spectrum is not necessary to be involved in material decomposition. In addition, realistic projection domain material decomposition has used a phantom calibration fashion instead of spectrum-involved method\cite{stenner2007,wu2016w}. With dual-energy material decomposition, this work aims to develop a segmentation-free indirect transmission measurement-type energy spectrum estimation method using model spectra and material images-based polychromatic forward projection\cite{zhao2016a}.

The paper is organized as follows. In Sec.~\ref{sec:method}, we describe our method for spectrum estimation and four major components of the method. In Sec.~\ref{sec:evaluation}, we describe evaluation studies which include numerical simulations, experimental phantom and realistic images. Section~\ref{sec:results} presents the results, and we conclude with discussion as well as summary in Sec.~\ref{sec:discussion}.

\section{METHODS}\label{sec:method}
To avoid determining each energy bin of the x-ray spectrum, we use model spectra to express the spectrum that is to be estimated. The model spectra expression can significantly reduce the degrees of freedom (DOFs) of the spectrum estimation problem. In this case, the unknown spectrum $\Omega(E)$ is the weighted summation of a set of model spectra $\Omega_{i}(E)$, i.e.,
\begin{equation}\label{equ:spek}
\Omega(E)=\sum_{i=1}^{M}c_{i}\Omega_{i}(E),
\end{equation}
with $M$ the number of the model spectra and $c_i$ the weight on the respective model spectrum. The model spectra can be predetermined using spectrum generators (such as Spektr~\cite{siewerdsen2004} and SpekCalc~\cite{poludniowski2009}) or MC simulation toolkits based on Geant4.

\begin{figure}[t]
	\begin{center}
	\begin{tabular}{c}
    \includegraphics[width=0.95\textwidth]{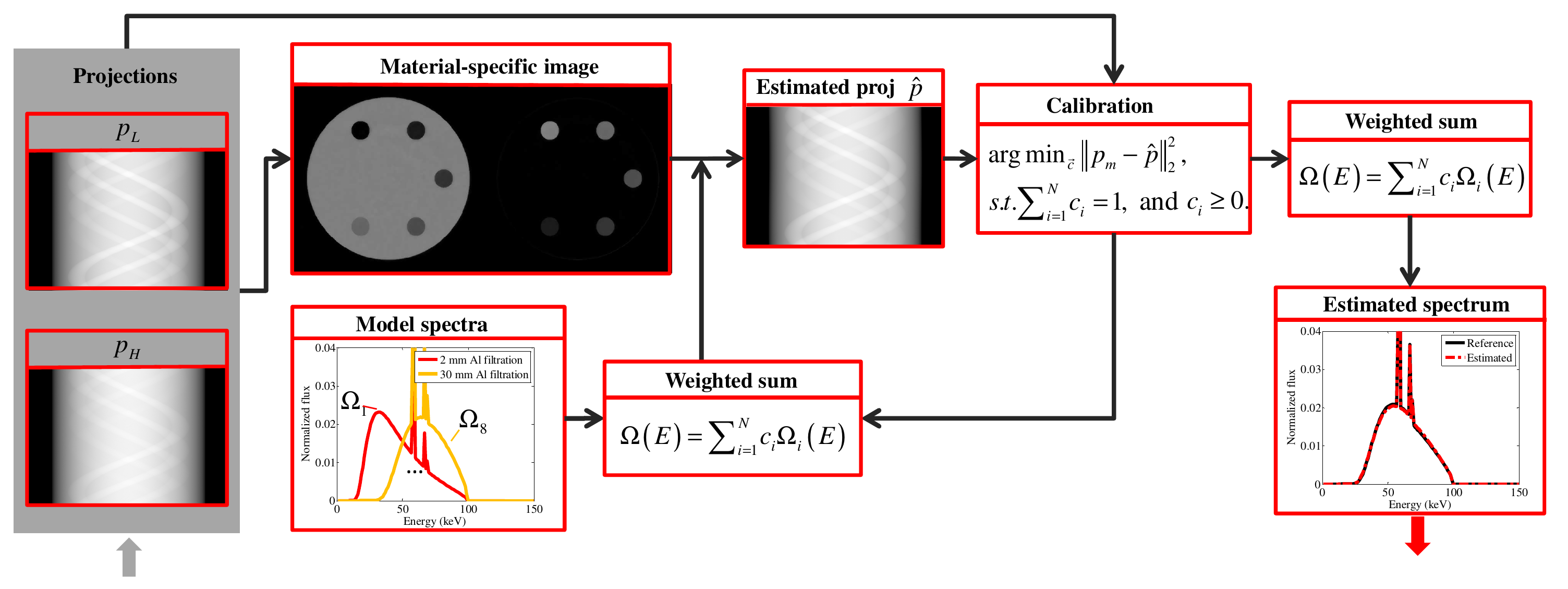}
    \end{tabular}
	\end{center}
    \caption{Flowchart of the proposed dual-energy material decomposition-based spectrum estimation method.}
    \label{fig:f1}
\end{figure}

The flowchart of the proposed algorithm is presented in Fig.~\ref{fig:f1}. The method starts from raw projection data. Material-specific images are then calculated by using dual-energy material decomposition algorithms. From the material images, along with the model spectra expression, a set of estimated projection $\hat{p}$ is calculated. By iteratively updating the unknown weights $c_i$, we can converge to a set of optimal $c_i$ to minimize the quadratic error between the measured raw projection $p_m$ and the estimated projection $\hat{p}$. The unknown spectrum is finally obtained by using Eq.~(\ref{equ:spek}). The four major components of the approach will be detailed in the following subsections: dual-energy material decomposition, material image-based polychromatic reprojection, weight estimation and model spectra generation.

\subsection{Dual-energy material decomposition}
\textcolor{black}{In order to obtain quantitative material-specific basis images, we use a nonlinear  empirical dual energy calibration (EDEC) algorithm to perform material decomposition\cite{stenner2007}. For the EDEC calibration, a two cylinder phantom with the same materials as the subsequent CT scan is used. The imaging protocol used in the calibration is the same as that used in the phantom studies. By minimizing the least square deviation between a set of basis images and a corresponding template, one can calibrate the decomposition coefficients for each basis material. }Since magnified noise is a general concern for both projection-domain and image-domain dual-energy material decomposition, in this study, to keep the accuracy of the estimated projection $\hat{p}$, we have \textcolor{black}{also} used an iterative image-domain method to obtain the noise significantly reduced material-specific images~\cite{niu2014}. The method optimizes an objective function which consists of a data fidelity term and a quadratic penalty term using the nonlinear conjugate gradient algorithm. \textcolor{black}{By using the regularization framework, the iterative image-domain material decomposition method provides noise well reduced while accuracy well preserved material-specific images.}

\subsection{Polychromatic projection on decomposed material images}
In dual-energy material decomposition, the linear attenuation coefficient $\mu(\vec{r},E)$ is modelled with two basis materials via a weighted summation fashion as,
\begin{equation}\label{equ:decomposition}
\mu(\vec{r},E)=f_{1}(\vec{r})\psi_{1}(E)+f_{2}(\vec{r})\psi_{2}(E).
\end{equation}
Here $\psi_{1,2}$ are the known independent energy dependencies which can be mass attenuation coefficients of basis materials and $f_{1,2}(\vec{r})$ are the material-selective images. Based on the above formulation, polychromatic projection of an object is represented as
\begin{equation}\label{equ:polyreprojBimg}
\hat{I}=N\int_{0}^{E_{max}}\mathrm{d}E\,\Omega(E) \, \eta(E)\,\mathrm{exp}\left[-A_{1}\psi_{1}(E)-A_{2}\psi_{2}(E)\right],
\end{equation}
with $A_{1}=\int_{L}\mathrm{d}\vec{r}\,f_{1}(\vec{r})$ and $A_{2}=\int_{L}\mathrm{d}\vec{r}\,f_{2}(\vec{r})$ the line integrals of the material-selective images. Here $L$, $\Omega(E)$ and $E_{max}$ are the propagation path length of each ray, the corresponding polychromatic x-ray spectrum of the ray and the maximum photon energy of the spectrum, respectively. $\eta(E)$ is the energy dependent response of the detector.
Note that $\hat{I}$ is detector pixel dependent and the detector channel index is omitted for convenience. 
After applying the logarithmic operation, the estimated projection data can be expressed as:
\begin{eqnarray}\label{equ:esproj}
\hat{p}(\vec{c})&=& \log\left(\frac{\hat{I}_{0}}{\hat{I}}\right) \\
&=&\log\left(\frac{\int_{0}^{E_{max}}\mathrm{d}E\,\Omega(E)\,\eta(E)}{\int_{0}^{E_{max}}\mathrm{d}E\,\Omega(E)\,\eta(E)\,f(A_1,A_2)}\right)\\
&=&\log\left(\frac{\sum_{i=1}^{M}c_{i}\int_{0}^{E_{max}}\mathrm{d}E\,\Omega_{i}(E)\,\eta(E)}{\sum_{i=1}^{M}c_{i}\int_{0}^{E_{max}}\mathrm{d}E\,\Omega_{i}(E)\,\eta(E)\,f(A_1,A_2)}\right),
\end{eqnarray}
where
\begin{equation}
f(A_1,A_2)=\mathrm{exp}\left[-A_{1}\psi_{1}(E)-A_{2}\psi_{2}(E)\right].
\end{equation}
Note that the air scan data $\hat{I}_{0}=N\int_{0}^{E_{max}}\mathrm{d}E\,\Omega(E) \, \eta(E)$ has been used in Eq.~(\ref{equ:esproj}) and $\hat{p}(\vec{c})$ is a function of unknown weights $\vec{c}$.

\subsection{Weights estimation}
To estimate the unknown weights for each model spectrum, we minimize the quadratic error between the detector measurement $p_m$ and the corresponding estimated projection $\hat{p}$ by iteratively updating the weights. This procedure is formulated as the following optimization problem,
\begin{equation}\label{equ:opt-constraint2}
\mathbf{c^*}=\underset{\mathbf{c}} {\mathrm{argmin}}\;\|p_{m}-\hat{p}(\vec{c})\|_{2}^{2},    ~~\mathrm{s.t.}~\sum_{i=1}^{M}c_{i}=1,~\mathrm{and}~c_{i}\geq0.
\end{equation}
Here the normalization constraint $\sum_{i=1}^{M}c_{i}=1$ and the non-negative constraint which keeps the solution of the problem physically \textcolor{black}{meaningful}, are introduced. The objective function has minimum when the spectrum expressed using the model spectra matches the unknown raw spectrum. To solve Eq.~(\ref{equ:opt-constraint2}), we use a sequential optimization approach, i.e., minimizing the objective function, followed by normalizing the solution and enforcing non-negative constraint sequentially. \textcolor{red}{To minimize the objective function, we have used a simple multi-variable downhill simplex method.} \textcolor{red}{The calibration procedure can be summarized as follows:}

\begin{adjustwidth}{0.6cm}{}
\textbf{Algorithm (Sequential optimization)}\\
1. Set k=0, choose $\mathbf{c}_0$.\\
2. \textbf{Repeat}\\
3. $\mathbf{c}^{k+1}=\underset{\mathbf{c}} {\mathrm{argmin}}\;\|p_{m}-\hat{p}(\mathbf{c}^k)\|_{2}^{2}$\\
4. $\mathbf{c}^{k+1}=\mathbf{c}^{k+1}/\mathrm{sum}(\mathbf{c}^{k+1})$\\
5. $\mathbf{c}^{k+1}=(\mathbf{c}^{k+1})_{+}$\\
6. $k\leftarrow k+1$\\
7. \textbf{Until} stopping criterion is satisfied.
\end{adjustwidth}


\subsection{Model Spectra generation}
There are several different ways to obtain the model spectra and they would affect the final results during the calibration procedure, thus well-validated spectrum generators which generate spectra with accurate physical properties (such as characteristic peak)~\cite{siewerdsen2004,poludniowski2009,Hernandez2014,hernandez2016,punnoose2016} need to be employed to calibrate the real spectrum. In the study, we have used MC simulation toolkit Geant4~\cite{allison2006} and the widely used software SpekCalc~\cite{poludniowski2009} to produce the model spectra. For both methods, different thicknesses of filters are added to yield model spectra with different half-value layers. \textcolor{black}{For the Geant4 model spectra simulation, we have used mono-energetic electrons (pencil beam) to hit the tungsten target. The thickness of the target is 0.25 cm and the target angle is 30 degree. For the SpekCalc model spectra, we have set the target angle to 30 degree and added a 0.8 mm Beryllium window. The energy bin width is 1 keV and air thickness is 1000 mm.}

\section{Evaluations} \label{sec:evaluation}
\subsection{Numerical study}
We first use numerical simulation to evaluate the proposed spectrum estimation method. A water cylinder with six iodine concentrate inserts (range 0-20 mg/mL with 4 mg/mL interval) was simulated in a 2D fan-beam CT geometry. The diameter of the water cylinder is 198 mm and the diameter of the six inserts are 22.5 mm. The low- and high-energy spectra are 100 kVp and 140 kVp, and they were generated using the SpekCalc software~\cite{poludniowski2009} with 12 mm Al and 0.4 mm Sn + 12 mm Al filtration, respectively. For the x-ray detection, an energy-integrating detector is simulated with 0.388 mm pixel size and 1024 pixels. The x-ray source to isocenter distance and to detector distance are 785 mm and 1200 mm, respectively. A set of 720 view angles were scanned in an angular range of $360^0$. Since one difficulty of DECT decomposition is the ill-conditioning, Poisson noise was introduced to raw projection to show robustness of the algorithm. In addition, first order beam hardening correction was performed to improve the accuracy of the material-specific images. During dual-energy material decomposition, regions-of-interest (ROIs) in the central water cylinder and in the 20 mg/mL iodine concentrate insert are used to calculate the decomposition matrix. 

To quantify the accuracy of the estimated spectrum, we calculate the normalized root mean square error (NRMSE) and the mean energy difference $\Delta E$ between the raw spectrum (ground truth) and the estimated spectrum, i.e.,

\begin{equation}\label{equ:error}
\mathrm{NRMSE}=\sqrt{\frac{\sum_{e=1}^{N}(\hat{\Omega}(e)-\Omega(e))^{2}}{\sum_{e=1}^{N}\Omega(e)^{2}}},    
\end{equation}
\begin{equation}\label{equ:meanEnergy}
\Delta E= \sum_{e=1}^{N}E_e\,(\Omega(e)-\hat{\Omega}(e)),
\end{equation}
with $\hat{\Omega}(e)$ the $e$th energy bin of the normalized estimated spectrum and $\Omega(e)$ $e$th energy bin of the normalized true spectrum. $N$ and $E(e)$ are the number of the energy bins and the energy of the $e$th energy bin of the spectrum, respectively.


For the numerical simulation study, the raw spectrum used to generate the projection data is obtained from SpekCalc software and the model spectra used to calibrate the raw spectra are also obtained from SpekCalc. In order to further cross check the robustness of the proposed method with respect to the model spectra, MC simulation toolkit Geant4 is also employed to generate the model spectra. Geant4 enables comprehensive physics modelling that embedded in a flexible structure and it offers a full list of electron interaction modelling which makes a important role in x-ray generation, including the bremsstrahlung effect and characteristic radiation. In addition, Geant4 collaboration has consistently released the G4EMLOW low energy package which enables access to precise cross-sections for x-ray photons production at very low energy scale.

In this study, eight model spectra are generated with different thickness of aluminum filtration with SpekCalc and Geant4. For the SpekCalc model spectra, the filtration is from 2 mm to 30 mm in step of 4 mm, while for the Geant4 model spectra, the filtration is from 8 mm to 22 mm in step of 2 mm. During each MC simulation, a total number of $5\times10^8$ electrons are emitted to hit the target.

\subsection{Comparison study}
\textcolor{black}{To further evaluate the performance of the proposed method, the method is compared to the previous segmentation-based ITM spectrum estimation method\cite{zhao2015}. Different from the ITM method, the proposed DECT-based method does not require a dedicated phantom calibration and image segmentation. To demonstrate this feature, we have performed spectrum estimation using a numerical anthropomorphic thorax phantom. In this study, in order to show the advantage of the DECT-based method, we have estimated the spectrum with both standard attenuation coefficients and intended mismatched attenuation coefficients. That is, raw projection data is first generated using standard NIST attenuation coefficients, and the data was then reconstructed to obtain CT images. For the segmentation-based method, the spectrum is estimated based on the segmented CT image which is assigned to standard coefficients. For the DECT-based method, the EDEC algorithm\cite{stenner2007} is employed to generate material images.}

\textcolor{black}{Since the attenuation coefficients of a realistic patient or object may deviate from the standard value, for example, a fatty body may have lower attenuation coefficient than the standard value, we then further estimate the spectrum using intended mismatched attenuation coefficients. In this case, raw projection data was generated using 98\% standard attenuation coefficients for both bone and tissue. Then both methods were employed to estimate the spectrum based on the raw projection data.}


\subsection{Experimental phantom study}
The algorithm was also evaluated using experimental phantom data acquired with a cone-beam CT (CBCT) benchtop system and an in-house rotating gantry small animal micro-CT scanner which was developed by the authors.

\subsubsection{Benchtop CBCT}
An anthrophomorphic head phantom was scanned using the CBCT benchtop system. The distance of source to isocenter (SOD) and source to detector (SDD) were 1000 mm and 1500 mm, respectively. A total of 655 projections were evenly acquired in 360 degree rotation with $2\times2$ rebinning mode and narrow collimation to avoid scatter radiation. Tube potentials of high and low-energy spectra were 125 kVp and 75 kVp, respectively. During CT data acquistion, \textcolor{black}{the x-ray source (Rad 94, Varian, Palo Alto, CA) and the flat detector (PaxScan 4030D, Varian, Palo Alto, CA) were stationary and the phantom was rotated to acquire projection from different view angles. The pixel matrix and size of the detector are 2048$\times$1536, 194 $\mu$m, respectively.} A prefiltering of 6.0 mm aluminum was always applied and the reference spectrum was simulated using SpekCalc with the filtration matched with the prefilter. Due to the large phantom size, scatter radiation would play an important role in projection. To reduce the impact of scatter radiation on the accuracy of estimated spectrum, we have acquired the data using narrow collimation.  

\subsubsection{Small animal micro-CT}

\begin{figure}
	\begin{center}
	\begin{tabular}{c}
    \includegraphics[width=0.6\textwidth]{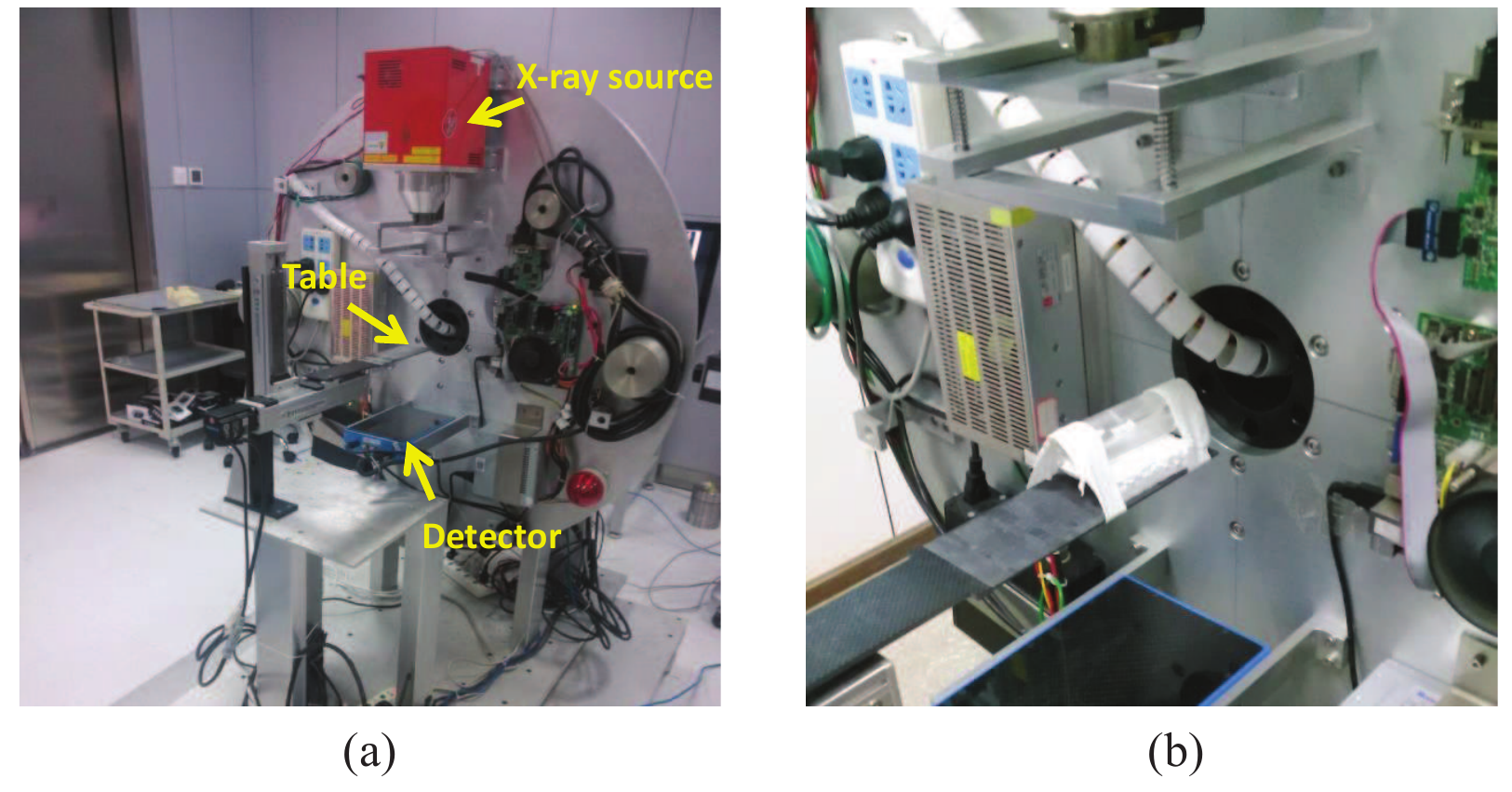}
    \end{tabular}
	\end{center}
    \caption{Experimental phantom study using the small animal CT scanner. (a) The system geometry for the CT scanner is rotating gantry, and the x-ray tube and detector are assembled on the gantry while the specimen lies stationary on the table between the x-ray source and the detector. (b) Phantom scanned using the micro-CT scanner.}
    \label{fig:microCT}
\end{figure}

A simplified mice phantom was scanned using a small animal micro-CT scanner which was developed by the authors in a gantry rotating geometry, as shown in Fig.~\ref{fig:microCT}. The x-ray source (L9421-02, Hammamatsu Photonics, Hammamatsu City, Japan) and detector (Dexela 1207, PerkinElmer, Waltham, MA) were assembled on the rotating gantry while the phantom was kept stationary during CT scan. \textcolor{black}{The pixel matrix and size of the detector are 1536$\times$864, 74.8 $\mu$m, respectively.} The SOD and SDD were 229 mm and 416 mm, respectively. A total of 450 projections were acquired for image reconstruction. To perform dual-energy CT imaging, the phantom was scanned with low-energy spectrum (40 kV with 2 mm aluminum filtration) and high-energy spectrum (90 kV with 0.3 mm copper filtration) in a sequential mode. For both phantom studies, low and high energy CT images were reconstructed by using filtered backprojection (FBP) algorithm.
~Low-energy data sets were used to estimate low-energy spectra and high-energy spectra estimation follows the same procedure.

\subsection{Realistic images evaluation}

To further evaluate the proposed method, we generated high- and low-energy projection data by forward projecting two sets of abdomen images which were acquired from a dual-source dual-energy CT scanner (Siemens SOMATOM Definition Flash, Siemens AG, Forchheim, Germany). The system enables high- and low-energy CT scans with two x-ray tubes and corresponding detectors mounted onto a single rotating gantry with $90^\circ$ angular offset. The x-ray tubes can operated independently with regard to tube filtration, tube voltage as well as tube current. Low- and high-energy projection data were obtained with 100 kVp and 140 kVp spectra, respectively. Images were reconstructed using the build-in commercial software.

Based on these CT images, an GPU-based forward projecting program was used to generate the high- and low-energy "realistic" abdomen projection images and these projection data sets were applied to spectra estimation. Iterative image-domain material decomposition method was used to yield noise reduced material-selective images. The iteration number was set to 200 and the control parameter $\beta$ which balances the data fidelity term and the regularization term, was set to 0.015. During the calibration procedure, eight model spectra were used. These spectra were generated using SpekCalc with aluminium filter from 6 mm to 20 mm with 2 mm interval. Note that in this evaluation, we did not have true spectrum, i.e., there was no ground truth for comparison. Thus we compared the forward projection data with the estimated projection $\hat{p}$ when the optimization problem was converged.

\section{RESULTS} \label{sec:results}

\subsection{Numerical simulation}

\begin{figure}
	\begin{center}
	\begin{tabular}{c}
    \includegraphics[width=0.6\textwidth]{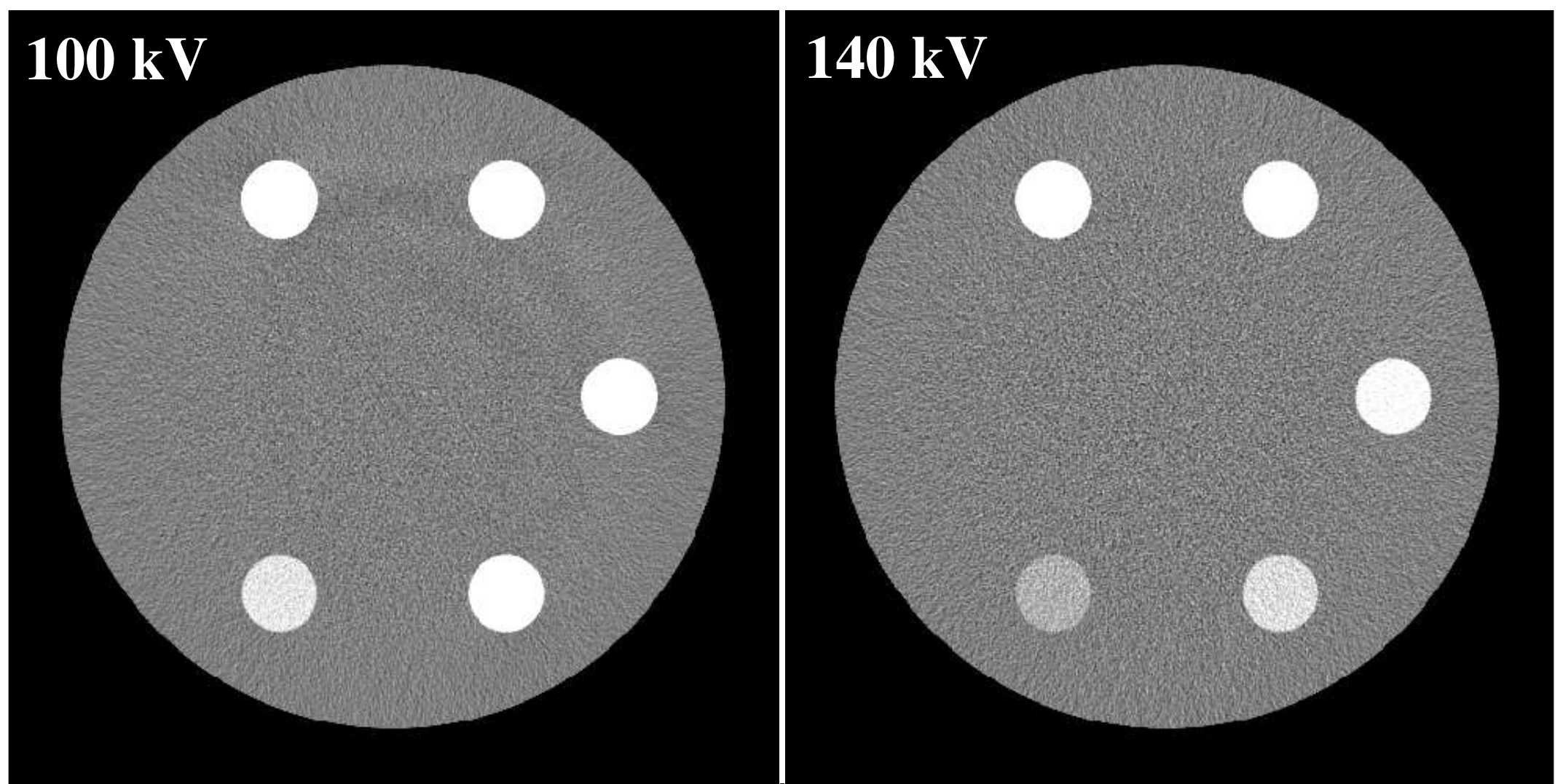}
    \end{tabular}
	\end{center}
    \caption{High and low-energy CT images (C/W = 0 HU/300 HU) of the numerical water phantom containing inserts with different iodine concentrates.  }
    \label{fig:f3}
\end{figure}

Figure~\ref{fig:f3} shows the results of low- and high-energy CT images of the numerical iodine concentrate phantom. As can be seen, 100 kV image shows much higher contrast level for the iodine inserts, as expected. During material decomposition, regions-of-interest in the central area of the numerical phantom and in the 20 mg/mL are used to calculate the decomposition matrix.

\begin{figure}[b]
	\begin{center}
	\begin{tabular}{c}
    \includegraphics[width=0.6\textwidth]{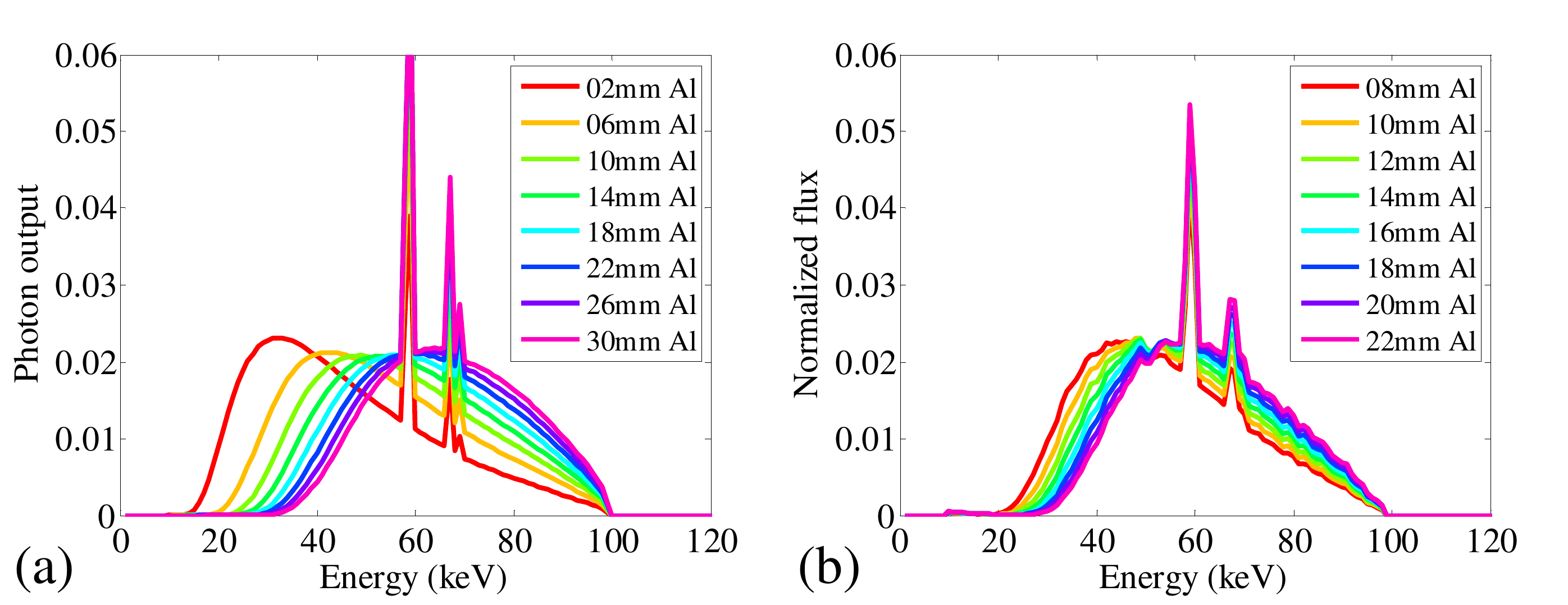}
    \end{tabular}
	\end{center}
    \caption{100 kVp model spectra generated using SpekCalc (a) and Monte Carlo toolkit Geant4 (b) with different thicknesses of aluminium filtration. Note that the noise level of the Geant4 model spectra is relatively high due to the limited 100 keV electron events.}
    \label{fig:f4}
\end{figure}

Model spectra used in the spectrum calibration procedure are shown in Fig.~\ref{fig:f4}. In order to evaluate the robustness of the proposed spectrum estimation method with respect to the model spectra generator, different methods are employed to obtain the model spectra. Figure~\ref{fig:f4}(a) shows the 100 kV model spectra generated using SpekCalc software with fix kVp setting and different thicknesses of filtration ranging from 2 mm to 30 mm in step of 4 mm. Figure~\ref{fig:f4}(b) shows model spectra simulated using MC toolkit Geant4 with filtration ranging from 8 mm to 22 mm in step of 2 mm. Due to the limited electron events emitted in the MC simulation, the Geant4 model spectra contain some noise which may impact the final result. For both model spectra, spectrum becomes harder and narrower as the thickness of the filtration increases.

\begin{figure}
	\begin{center}
	\begin{tabular}{c}
    \includegraphics[width=0.6\textwidth]{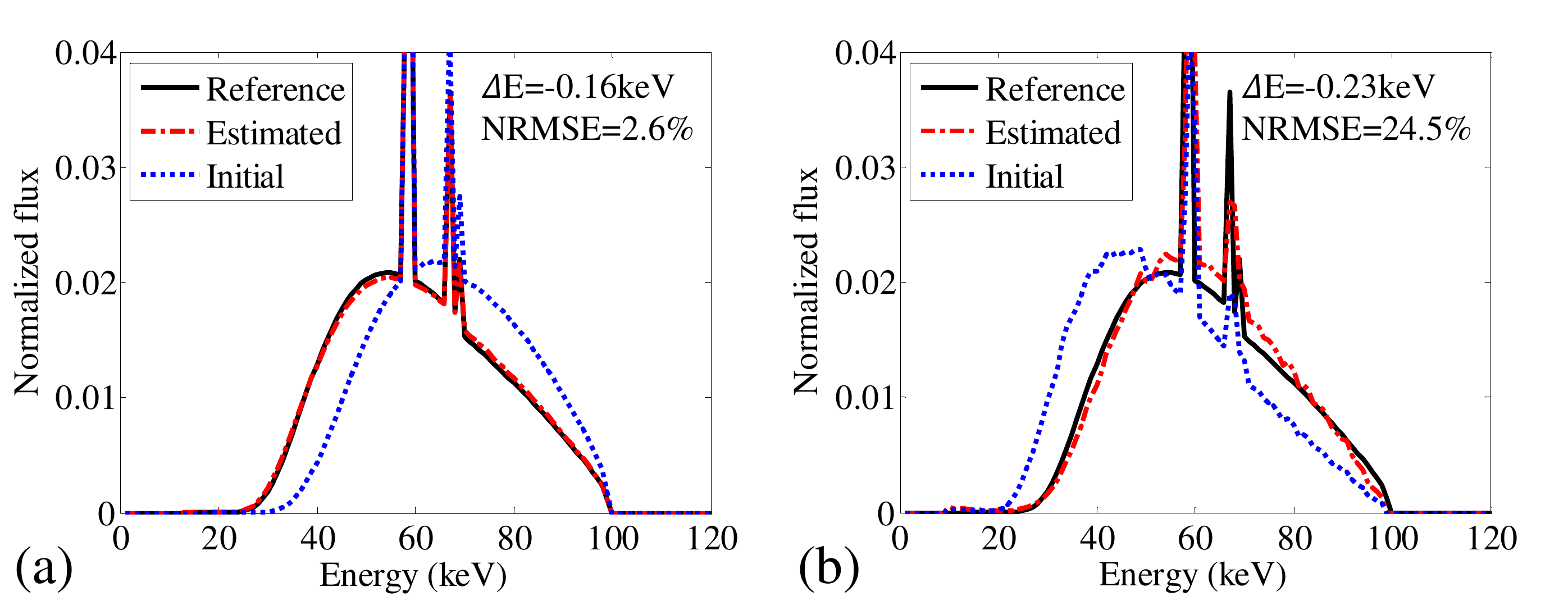}
    \end{tabular}
	\end{center}
    \caption{X-ray spectra estimated for the numerical iodine concentrate phantom data. Model spectra generated using both SpekCalc (a) and Geant4 (b) were employed in the estimation procedure. The initial guesses for the sequential optimization problem correspond to the hardest and softest model spectra for the SpekCalc and Geant4 model spectra estimation, respectively. Since the Geant4 model spectra are relatively noisy, spectrum estimated used these spectra is also noisy. }
    \label{fig:f5}
\end{figure}

Figure~\ref{fig:f5} depicts the results of the 100 kV spectrum estimation for the numerical phantom data using both SpekCalc and Geant4 model spectra. The initial spectrum to calibrate the unknown weights is the hardest and softest model spectra for the SpekCalc and Geant4 model spectra estimation, respectively. The raw spectrum is the spectrum that was used to generate the 100 kV projection data and it can be regarded as the reference spectrum. The estimated spectrum matches the raw spectrum quite well when SpekCalc model spectra are used (as illustrated in Fig.~\ref{fig:f5}(a)), and $\Delta E$ and NRMSE are -0.16 keV and $2.6\%$, respectively, suggesting the dual-energy material decomposition-based method provides accurate spectrum estimate.

For the numerical simulation studies, since the raw projection is produced by the polychromatic forward projection using SpekCalc spectrum, spectrum estimation using different model spectra is also conduct, which could be a cross-check for the proposed method. Figure~\ref{fig:f5}(b) shows spectrum estimated using Geant4 model spectra and there are some noise in the estimated spectrum. These noises come from the Geant4 model spectra, as shown in Fig.~\ref{fig:f4}(b). Although the presence of noises cause an significantly increased NRMSE between the estimated spectrum and the true spectrum, their mean energy difference $\Delta E$ is -0.23 keV which is still comparable to that of spectrum estimation using SpekCalc model spectra.

\begin{figure}
	\begin{center}
	\begin{tabular}{c}
    \includegraphics[width=0.5\textwidth]{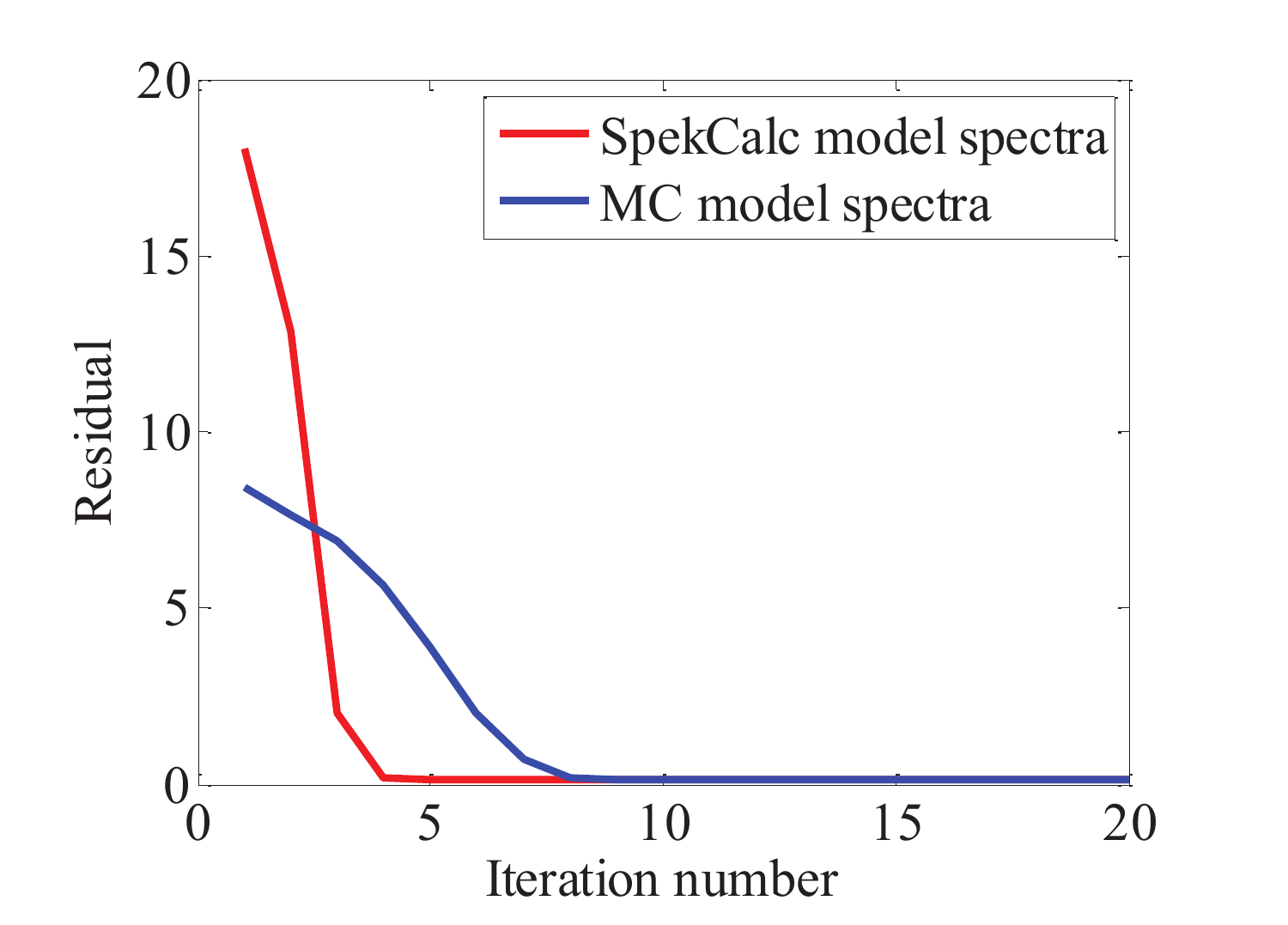}
    \end{tabular}
	\end{center}
    \caption{Residual between the raw projection and the estimated projection as the iteration number increases for the numerical iodine concentrate phantom data. }
    \label{fig:f6}
\end{figure}

Although the model spectra have effect on the final result, the proposed method tends to yield an optimal spectrum that minimize the quadratic error of the raw projection data and the estimated reprojection data by using the different model spectra. Figure~\ref{fig:f6} shows the residual between the raw projection data and the estimated projection as the iteration number of the optimization problem increases for the numerical simulation. For both model spectra, the objective function (Eq.~(\ref{equ:opt-constraint2})) converges to the same level, indicating the optimization procedure is robust against the model spectra. This is the reason why the mean energy different can keep at the same level for different model spectra.

\begin{figure}
	\begin{center}
	\begin{tabular}{c}
    \includegraphics[width=0.8\textwidth]{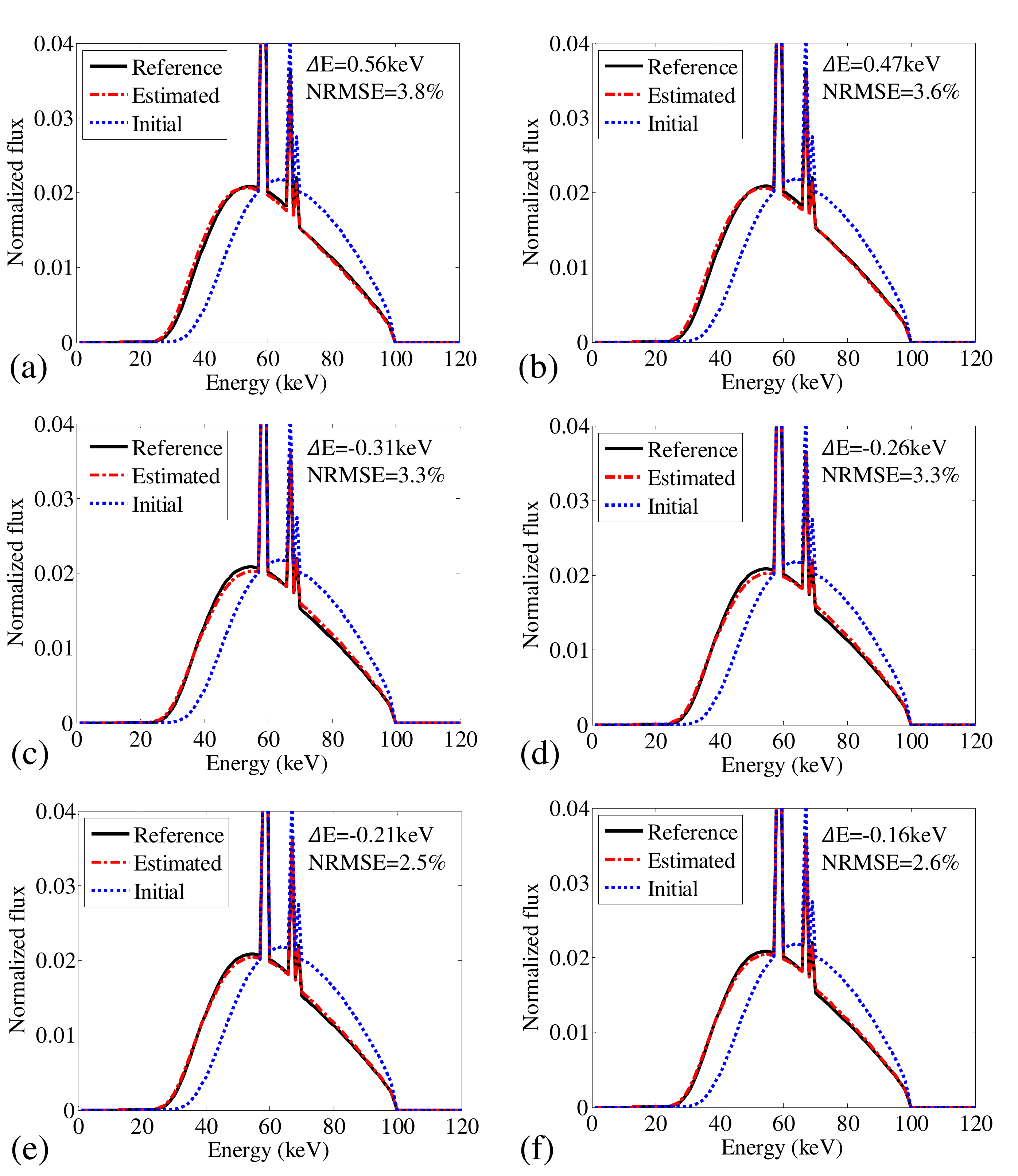}
    \end{tabular}
	\end{center}
    \caption{X-ray spectra estimated from the numerical phantom at different number of primary photon events. (a, c, e) use direct matrix inversion to obtain the material images for the spectrum estimation, while (b, d, f) use noise reduced dual-energy material decomposition method (i.e. iterative image-domain method) for the spectrum estimation. The numbers of the primary photon events used in the numerical simulations of the low-energy CT scans for (a, b), (c, d) and (e, f) are $3\times10^3$, $3\times10^4$ and $3\times10^5$, respectively. For each of the dual-energy CT scans, the numbers of the photon events of the high-energy CT scans are half of the respective low-energy CT scans. }
    \label{fig:f7}
\end{figure}

Since dual-energy material decomposition would amplify the image noise which may affect the final result, spectrum estimation using different numbers of primary photon events (i.e. dose levels) are performed to demonstrate the robustness of the proposed method. The first, second, and third row of Fig.~\ref{fig:f7} depict x-ray spectra estimated from the numerical phantom data by using $3\times10^3$, $3\times10^4$ and $3\times10^5$ photon histories, respectively. While the first column of Fig.~\ref{fig:f7} shows spectrum estimated using material-specific images obtained from direct matrix inversion (i.e., without noise reduction), the second column of Fig.~\ref{fig:f7} shows spectrum estimated using material-specific images obtained from iterative image-domain material decomposition method (i.e., with noise reduction). As can be seen, for different noise levels, the proposed method can accurately recover the true spectrum. For different dose levels, the maximum $\Delta E$ is 0.56 keV and the maximum NRMSE is $3.8\%$.

\subsection{Comparison studies}
\begin{figure}
	\begin{center}
	\begin{tabular}{c}
    \includegraphics[width=0.8\textwidth]{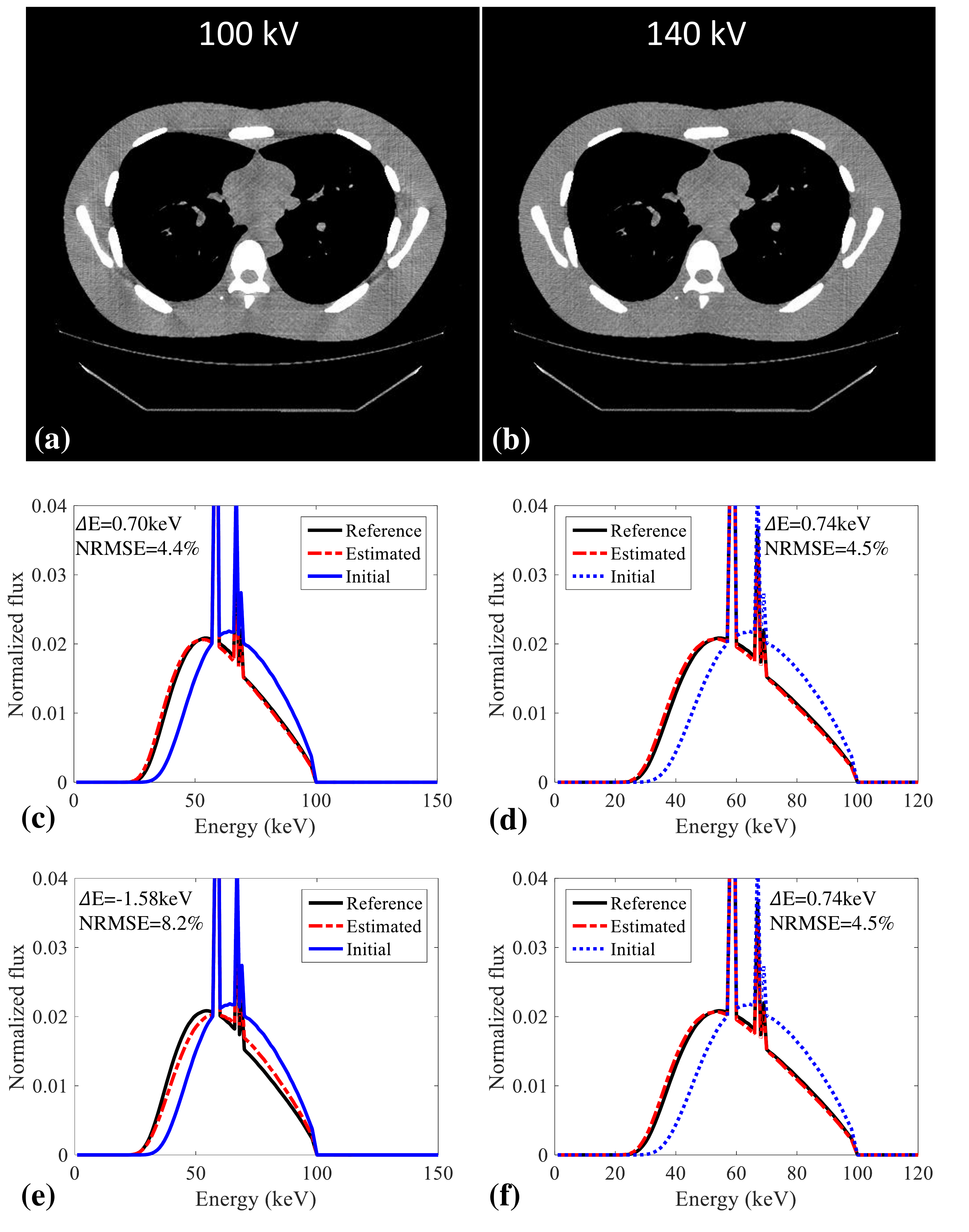}
    \end{tabular}
	\end{center}
    \caption{Results of energy spectrum estimation using the anthropomorphic thorax phantom. (a) and (b) are the 100 kV and 140 kV CT images, respectively. Display window for the images: [-150HU, 150HU]. (c) and (d) are the segmentation-based and the proposed DECT-based spectra estimation using raw projection data acquired with standard attenuation coefficients, respectively. (e) and (f) are the segmentation-based and the DECT-based spectra estimation using raw projection data acquired with 98\% attenuation coefficients. }
    \label{fig:f7_1}
\end{figure}

\textcolor{black}{Figure~\ref{fig:f7_1} shows the results of spectra estimation using the anthropomorphic thorax phantom. Figure~\ref{fig:f7_1}(a) and (b) are the polychromatic 100 kV and 140 kV CT images, respectively. Compared to the 140 kV image, there are more beam hardening artifacts in the 100 kV image.  Figure~\ref{fig:f7_1}(c) and (d) show the segmentation-based and the proposed DECT-based spectra estimation using raw projection data obtained with standard attenuation coefficients. As can be seen, both methods can accurately estimate the spectrum. The NRMSE between the estimated spectra and their reference for the segmentation-based method and the DECT-based method are 4.4\% and 4.5\%, respectively. Figure~\ref{fig:f7_1}(e) and (f) show the segmentation-based and the DECT-based spectra estimation using raw projection data obtained with 98\% standard attenuation coefficients. As can be seen, for the non-standard material, the NRMSE of the estimated spectrum using segmentation-based method is increased from 4.4\% to 8.2\%. However, with the DECT-basd method, the accuracy of the estimated spectrum is well preserved. }

\subsection{Experimental phantom studies}

\begin{figure}[t]
	\begin{center}
	\begin{tabular}{c}
    \includegraphics[width=0.6\textwidth]{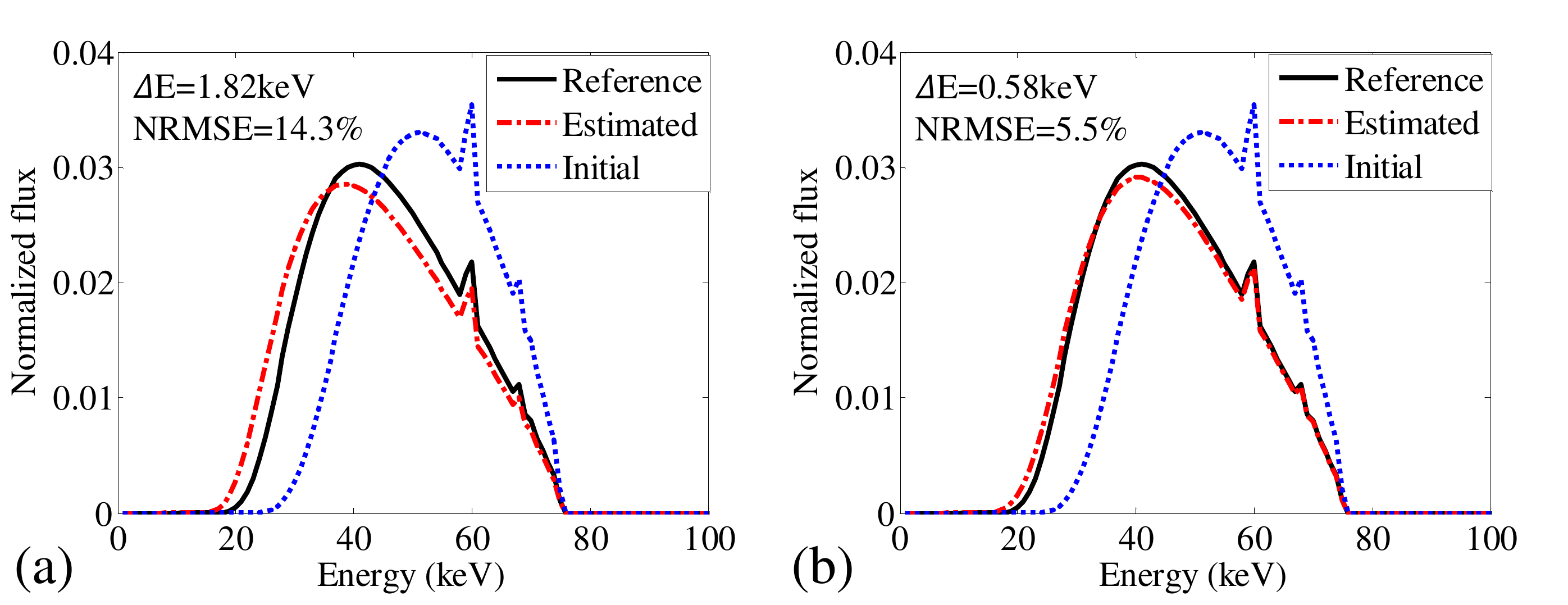}
    \end{tabular}
	\end{center}
    \caption{Low- and high-energy CT images of the experimental head phantom. Display window: [-300 HU, 300 HU].}
    \label{fig:f9}
\end{figure}

\begin{figure}[b]
	\begin{center}
	\begin{tabular}{c}
    \includegraphics[width=0.8\textwidth]{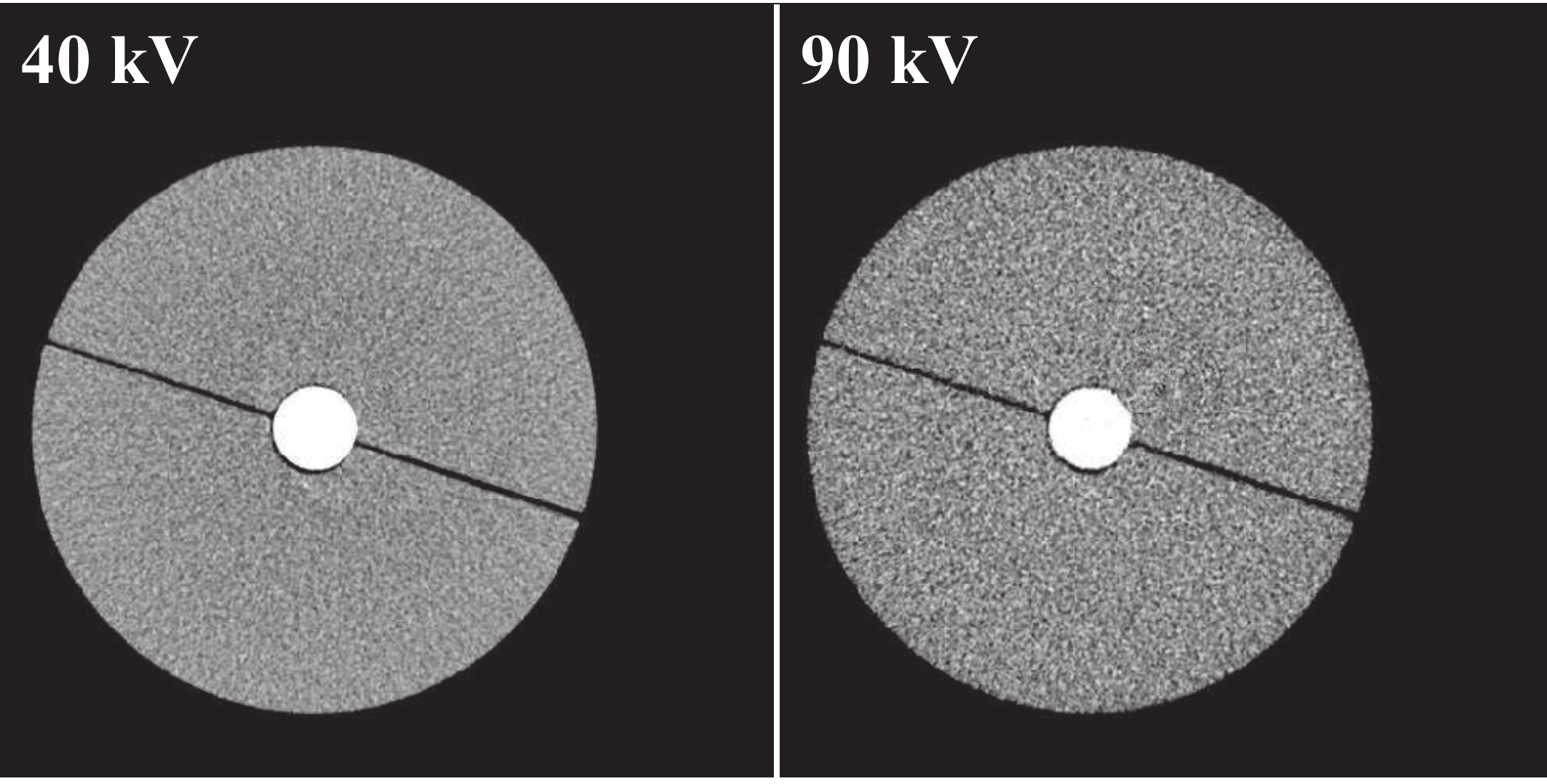}
    \end{tabular}
	\end{center}
    \caption{Spectra estimated using the physical head phantom with (b) and without (a) detector efficiency incorporation.}
    \label{fig:f10}
\end{figure}

Figure~\ref{fig:f9} shows low- and high-energy CT images of the experimental head phantom. Figure~\ref{fig:f10} depicts spectrum estimated with the anthrophomorphic head phantom with and without detector efficiency incorporation. For this experimental evaluation, the benchtop CBCT system has used a flat detector with 0.6 mm thickness of CsI. To better estimate the spectrum, energy dependent efficiency should be taken into account. $\Delta E$ is reduced from 1.82 keV and 0.58 keV after detector efficiency incorporation, while NRMSE is reduced from $14.3\%$ to $5.5\%$ after detector efficiency incorporation. For both cases, the initial spectra for the constrained optimization problem are the hardest model spectra. The reference true spectrum is generated using SpekCalc with filtration matches with the filtration that used in the experimental data acquisition.


\begin{figure}
	\begin{center}
	\begin{tabular}{c}
    \includegraphics[width=0.6\textwidth]{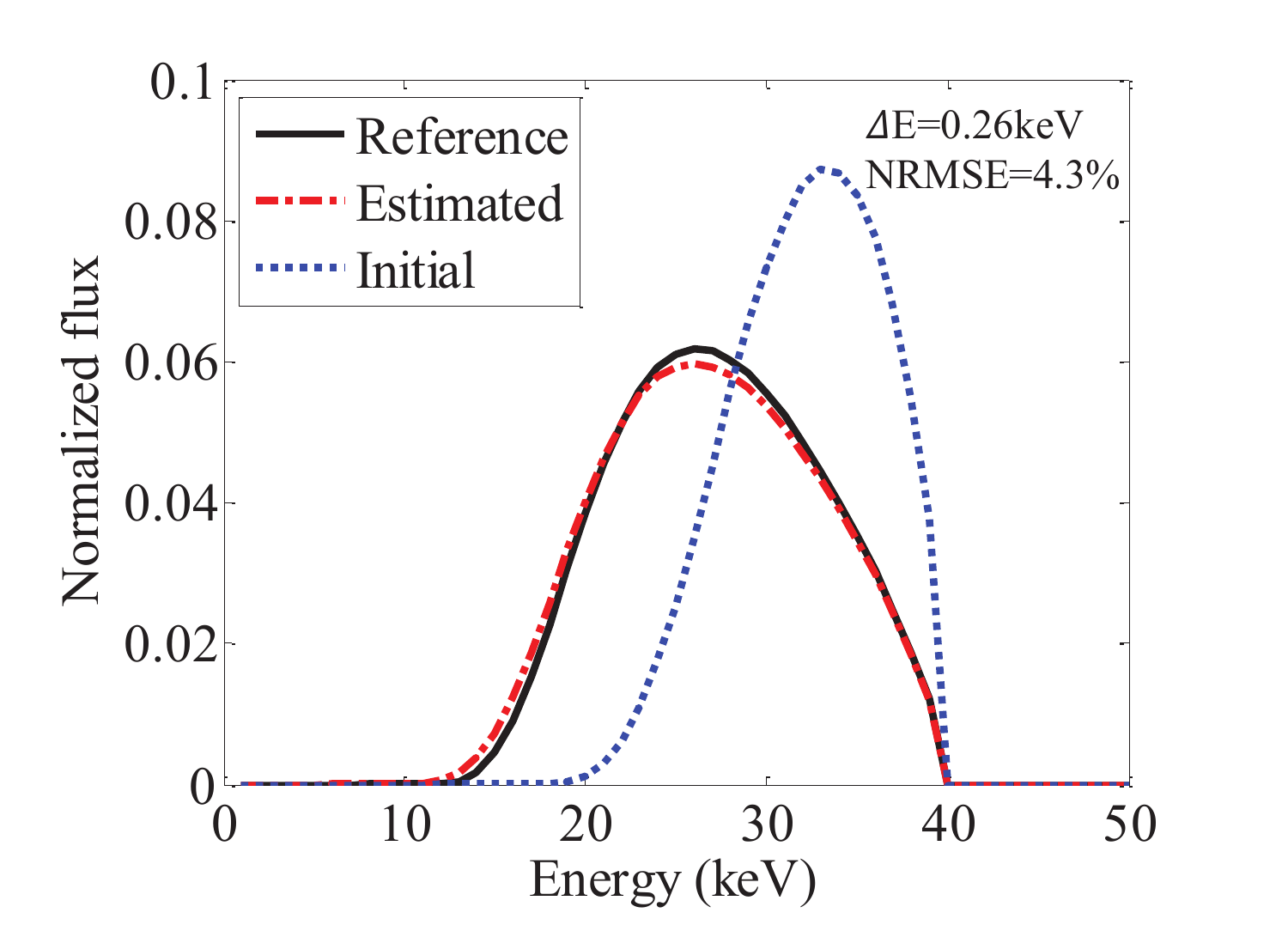}
    \end{tabular}
	\end{center}
    \caption{Low- and high-energy CT images of the mice phantom scanned using the micro-CT scanner. Display window: [-600 HU, 600 HU].}
    \label{fig:f9-2}
\end{figure}

\begin{figure}[t]
	\begin{center}
	\begin{tabular}{c}
    \includegraphics[width=0.5\textwidth]{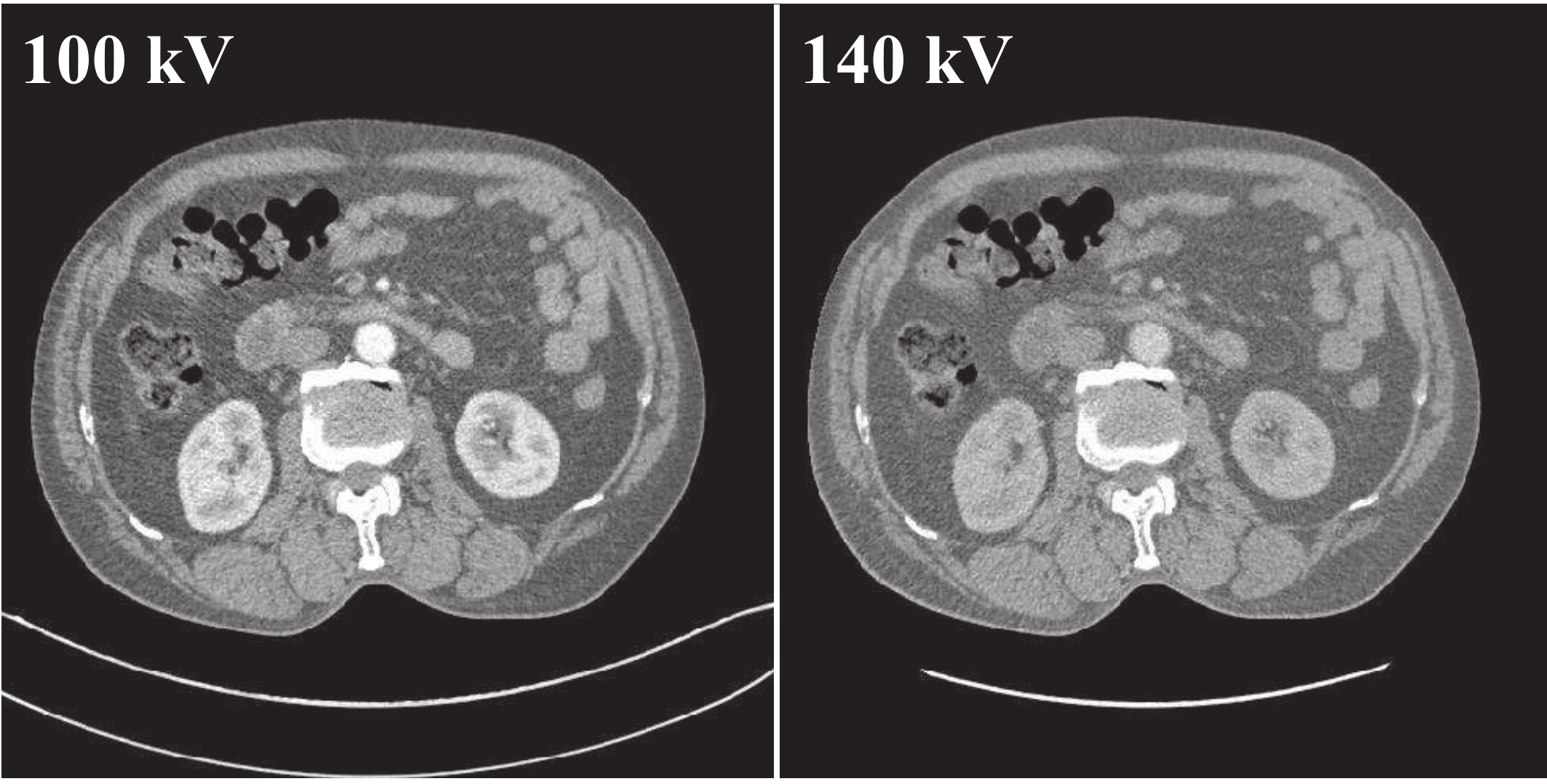}
    \end{tabular}
	\end{center}
    \caption{Spectra estimated using the mice phantom. The initial spectrum for the optimization problem is the hardest model spectra.}
    \label{fig:f10-2}
\end{figure}

Figure~\ref{fig:f9-2} shows low- and high-energy CT images of the mice phantom scanned using the in-house micro-CT scanner. Figure~\ref{fig:f10-2} depicts estimated low-energy spectrum. The initial spectrum for the constrained optimization problem is the hardest model spectra. The reference true spectrum is generated using SpekCalc with filtration matches with the filtration that used in the experimental data acquisition.


\subsection{Realistic images evaluations}

\begin{figure}
	\begin{center}
	\begin{tabular}{c}
    \includegraphics[width=0.6\textwidth]{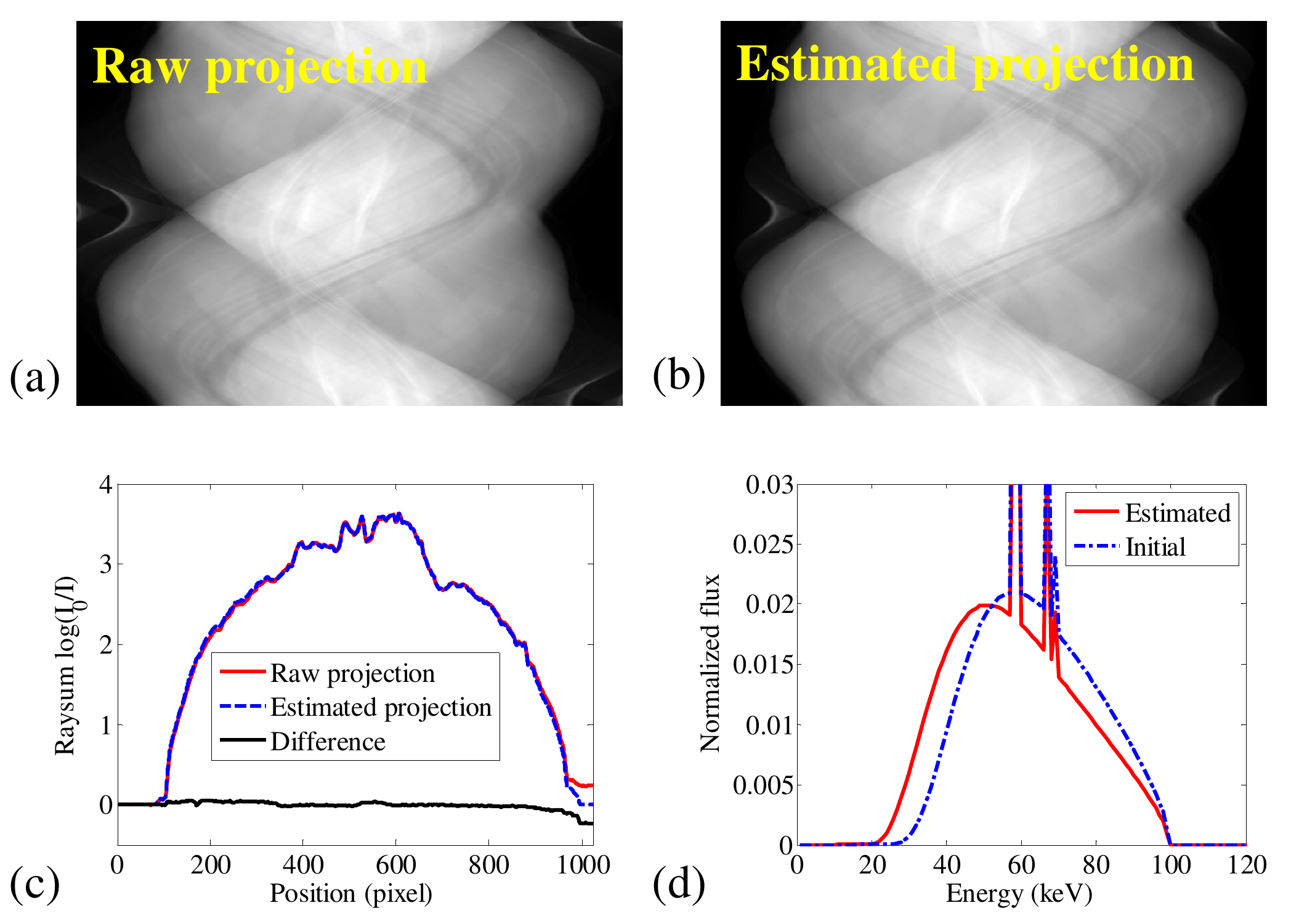}
    \end{tabular}
	\end{center}
    \caption{Axial contrast-enhanced, dual-energy abdominal CT scan using Siemens SOMATOM Definition Flash CT scanner. Display window: [-300 HU, 300 HU].}
    \label{fig:f11}
\end{figure}

To confirm the results of the numerical and experimental phantom studies, we used ¡°realistic¡± patient data obtained by forward projecting two abdomen CT images. These two CT images (as illustrated in Fig.~\ref{fig:f11}) were acquired and reconstructed with a Siemens SOMATOM Definiton Flash scanner under 100 kV and 140 kV. The forward projection data sets are used to mimic dual-energy low- and high-energy raw projection data.

\begin{figure}
	\begin{center}
	\begin{tabular}{c}
    \includegraphics[width=0.8\textwidth]{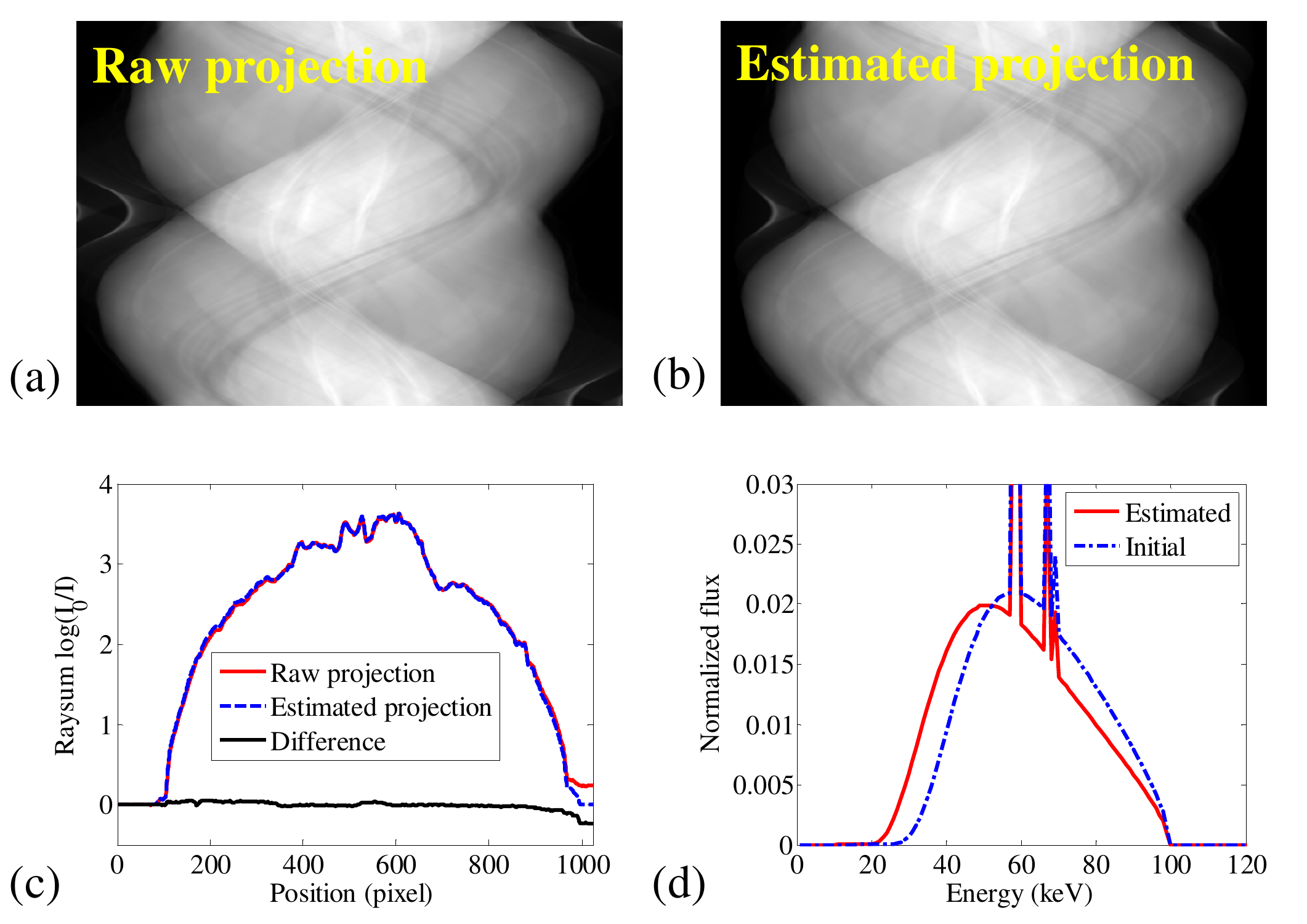}
    \end{tabular}
	\end{center}
    \caption{Results for realistic abdomen images evaluation. (a) Raw projection obtained by forward projecting the 100 kV CT image. (b) Estimated projection obtained using the estimated 100 kV spectrum. (c) Line profiles comparison of the raw projection and the estimated projection. (d) Estimated spectrum and the initial spectrum used in the optimization problem. Note that true spectrum is not available in this case and the estimated spectrum should be an effective spectrum which takes the system configuration and postprocessing algorithms into account.   }
    \label{fig:f13}
\end{figure}

During dual-energy material decomposition, adipose and iodine are chosen as the basis materials. For this evaluation, the true spectrum which is employed in raw projection data acquisition is not available, therefore, we compare raw projection data with the estimated projection which is calculated using the estimated spectrum when the objective function is converged. Figure~\ref{fig:f13} shows raw projection, estimated projection, their line profiles and the estimated spectrum. The final estimated projection matches with the raw projection quite well, indicating the proposed method can be applied to realistic cases.

\section{Discussion}
\label{sec:discussion}

In this study, we present a polychromatic x-ray spectrum estimation method based on dual-energy material decomposition. Different from the method proposed in Ref.~\citenum{zhao2015} where a segmentation procedure is needed for polychromatic reprojection, here, polychromatic reprojection is performed on material-specific images which were obtained by dual-energy material decomposition. This enables the proposed method to be segmentation-free. \textcolor{black}{More importantly, based on the DECT material decomposition, the proposed \textcolor{red}{method} can be further applied to phantom or patient data whose attenuation coefficients are unknown. This enables the proposed DECT-based technique to be applied to more realistic and general applications. Namely, the proposed method does not require a dedicated phantom calibration. Hence, the method can be regarded as a major step toward spectrum estimation of realistic applications without breaking the current workflow.}

\textcolor{black}{In realistic applications, x-ray spectrum is affected by many parameters, such as focal spot size and thickness of the target. In addition, the incident electrons used for hitting the target are also polychromatic. Thus, it's almost impossible to model all of the parameters and physical effects to obtain the true spectrum. Instead, the proposed method uses model spectra to span a space and the spectrum calibration procedure is to find an optimal spectrum in the space to generate a reprojection dataset whose difference is as small as possible from the raw projection data, i.e., the optimal spectrum has the most similar attenuation property as the true spectrum.}
Model spectra should affect the final results, as indicated in Fig.~\ref{fig:f5}. However, no matter what model spectra are used, the proposed method tends to yield a spectrum that minimize the quadratic error between the raw projection and estimated projection, as characterized by the objective function of Eq.~(\ref{equ:opt-constraint2}). To our belief, the NRMSE could be significantly reduced if the Geant4 model spectra contain less noises in the numerical simulation study. Nevertheless, some widely used and well validated spectrum generators including SpekCalc and Spektr are suggested to generate model spectra for this method. \textcolor{red}{For the weights calibration, since no automatic exposure control is used the numerical studies and experimental phantom study, we only use projection in one view angle to estimate the spectrum and the calibration time is less than 5 seconds on a personal PC (Intel Core i7-6700K CPU).}

To reduce the degrees of freedom of the spectrum estimation problem, we have employed a linear model (i.e. the unknown spectrum is expressed as a linear combination of a set of model spectra). However, it is unnecessary to use the linear model (i.e. weighted summation of a set of model spectra) and other models like a non-linear model can also be used to express the unknown spectrum. In this case, a simple curve and its widening or deforming versions can be used as model spectra. To better recover the unknown, we should also take advantage of the prior knowledge of an x-ray spectrum, namely, an x-ray spectrum contains characteristic x-rays and bremsstrahlung. Hence, some constraints should be added to the simple curve and its transformations. This is the reason why we have used filtered model spectra to fit the true spectrum in this study and these spectra can also be regarded as transformations (deformed by a polychromatic Beer's law) of a simple polychromatic x-ray spectrum. In future study, we would like to estimate the unknown spectrum using model spectra derived from a simple spectrum with the same kVp as the unknown spectrum and its high-order transformations (such as filtered using different materials).

For the experimental phantom study, the discrepancy between the estimated spectrum and the raw spectrum is much larger than than that in the simulation study. The discrepancy can be attributed to the  following three aspects : (1) Scatter is not considered in the reprojection procedure, thus the inevitable concomitant scatter radiation in the raw projection data would affect the accuracy of the spectrum result. To further refine the result, one may want to perform scatter correction~\cite{zhu2009scatter} before spectrum estimation. (2) For the material images-based polychromatic forward projection, the attenuation coefficients of the materials (bone and tissue) were obtained from the NIST database. These values may deviate from the real values of the head phantom. (3) The detector housing and sensor protection material would also affect the accuracy of the estimated spectrum.

\textcolor{black}{For the realistic patient study, we do not have access to the raw projection data on real scanners. In this case, we only have CT images and the images usually have performed water correction. If one simply forward projects the corrected CT images and then uses the projection to estimate spectrum, the resulted spectrum should be an effective spectrum which has taken the contribution of beam hardening procedure into account. }

\section{Conclusions}    \label{sec:conclusion}
This work presents an x-ray energy spectrum calibration method for CT scanners using dual-energy material decomposition and the indirect transmission measurement framework. The method performs polychromatic reprojection using material-specific images instead of segmented CT images, with which the segmentation procedure is avoided.  The reprojection data is then compared to the raw projection data and their difference is minimized by iteratively updating a set of weights, which are used to express the unknown spectrum together with a set of model spectra. The method was evaluated using numerical simulation data, experimental phantom data and realistic patient data. The results demonstrate raw spectra can be accurately recovered by incorporating the energy-dependent detector absorption efficiency. \textcolor{black}{The method does not require dedicated phantom calibration or knowledge of the material attenuation coefficient. It is promising for spectrum estimation of realistic applications.}



\subsection*{Disclosures}
No conflicts of interest to declare.

\acknowledgments
This work is supported by the Zhejiang Provincial Natural Science Foundation of China (Grant No. LR16F010001), National High-tech R\&D Program for Young Scientists by the Ministry of Science and Technology of China (Grant No. 2015AA020917), National Key Research Plan by the Ministry of Science and Technology of China (Grant No. 2016YFC0104507), Natural Science Foundation of China (NSFC Grant No. 81201091, 61601190, 51305257) and National Institute of Health (NIH 1R01 EB016777).


\bibliographystyle{spiejour}   


\vspace{2ex}\noindent\textbf{Wei Zhao} received his PhD in particle physics and nuclear physics from the Institute of High Energy Physics, Chinese Academy of Sciences in 2012. He is currently a research scientist in the Department of Radiation Oncology at Stanford University. His research interest is CT system development, X-ray imaging methods and devices. He is an NSFC-funded investigator.

\vspace{2ex}\noindent\textbf{Lei Xing} received his PhD from the the Johns Hopkins University in 1992. He is currently the Jacob Haimson Professor of Medical Physics and Director of Medical Physics Division of Radiation Oncology Department at Stanford University. His research has been focused on inverse treatment planning, tomographic image reconstruction, CT, optical and PET imaging instrumentations, image guided interventions, nanomedicine, and applications of molecular imaging in radiation oncology. He is an NIH, DOD, NSF, ACS-funded investigator and is active in numerous professional organizations. He is a fellow of the AAPM and AIMBE.

\vspace{2ex}\noindent\textbf{Qiude Zhang} received his BS degree in 2013 and his MS degree in biomedical engineering in 2016 from Huazhong University of Science and Technology. His research interest is CT system and X-ray detector development.

\vspace{2ex}\noindent\textbf{Qingguo Xie} received his PhD degree in control theory and control engineering from Huazhong University of Science and Technology (HUST) in 2001. He is currently a professor of biomedical engineering at HUST, where he directs the digital PET Lab. His research interests include system development and clinical application of positron emission tomography (PET). He is an NSFC-funded investigator.

\vspace{2ex}\noindent\textbf{Tianye Niu} received his BS degree in modern physics and his PhD in physical electronics from the University of Science and Technology of China in 2008. He is currently a professor of biomedical engineering in the School of Medicine at the Zhejiang University. His research interests include system development and clinical application of CT. He is an NSFC-funded investigator.


\listoffigures

\end{spacing}
\end{document}